\documentclass[prl,aps,floats,twocolumn,showpacs]{revtex4}
\usepackage[dvips]{epsfig}

\begin{document}

\def\sqr#1#2{{\vcenter{\hrule height.#2pt
   \hbox{\vrule width.#2pt height#1pt \kern#1pt
      \vrule width.#2pt}
   \hrule height.#2pt}}}
\def\square{{\mathchoice\sqr64\sqr64\sqr{3.0}3\sqr{3.0}3}}

\title{Two exercises about neutrino departure times at CERN}

\author{Bernd A. Berg and Peter Hoeflich}

\affiliation{Department of Physics, Florida State
             University, Tallahassee, FL 32306-4350, USA}

\date{\today } 

\begin{abstract}
Two simple exercises are solved, which educators can use to awake 
interest of their students in subtleties of the CERN Neutrino beam
to Grand Sasso (CNGS) experiment. 
The first one is about the statistical error of the average departure 
time of neutrinos from CERN.
The second one about a hypothetical bias in the departure times.

\end{abstract}
\maketitle


In a highly publicized measurement \cite{CNGS} neutrinos from CERN
arrived at Gran Sasso about 
\begin{eqnarray} \label{deviation}
  \delta t = 60\, {\rm ns\ (nano\ seconds)}
\end{eqnarray}
too early. Here we perform two pedagogical exercises using numbers 
given in the paper. 

The neutrinos are produced in extractions that last about 10,500~ns 
each, smashing protons into a graphite neutrino production target. 
About $6.2\times 10^{15}$ protons participate in one extraction and 
any proton could have produced the finally observed neutrino. 
Altogether 16,111 neutrino events entered the analysis.
\medskip

{\bf Exercise 1:} Assume the proton distribution is uniform in time,
what is the initial uncertainty in the neutrino creation time?

This is easy, let us center the time interval about zero, i.e., 
from $[-t_{\max},t_{\max}]$ with $t_{\max}=5,250\,{\rm ns}$. The
variance about zero is then
\begin{eqnarray} \label{var_t}
  \sigma^2(t)\ =\ \frac{1}{2\,t_{\max}} \int_{-t_{\max}}^{t_{\max}} 
  t'^{\,2}\,dt'\ =\ \frac{t_{\max}^2}{3}\,.
\end{eqnarray}
The error bar for the creation time becomes 
\begin{eqnarray} \label{eb_t}
  \triangle t\ =\ \sqrt{\frac{\sigma^2(t)}{16,111}}\,{\rm ns}\
  =\ 23.9\,{\rm ns}\,.
\end{eqnarray}
This is considerably larger than the final statistical error bar of 
6.7~ns given in \cite{CNGS}. As seen from Figs.~4 and~11 of the 
reference, the explanation appears to be: The true proton distribution 
is not uniform and drops off at the beginning and end of the time 
interval.  For our simple exercise we like the proton distribution to 
be flat, but let us use a new $t_{\max}$ so that $\triangle t$ becomes 
$\triangle t = 6.7\,{\rm ns}$. Solving (\ref{eb_t}) for $t_{\max}$ 
gives $t_{\max} =1,473\,{\rm ns}$, which is small on the scale of the
figures, where $2\,t_{\max}$ should cover about 10,000~ns. Using $t_{\max} 
= 5,000\,{\rm ns}$ gives still
\begin{eqnarray} \label{eb2_t}
  \triangle t\ =\ 22.7\,{\rm ns}\,.
\end{eqnarray}
To explain the discrepancy with the error bar of the reference to 
students, one has to take into account the structure and statistical 
significance of the data and/or analyze individual events at the 
boundaries. This requires access to the complete data sets underlying 
Fig.~11.
\medskip

{\bf Exercise 2:} Let us give up the assumption that each proton
creates neutrinos with the same probability. Instead, let us assume
that each proton diminishes by a factor 
\begin{eqnarray} \label{f}
  0\ <\ f\ =\ 1-p\ <\ 1
\end{eqnarray}
just a little bit the ability of the subsequent protons to create 
neutrinos. Let us assume that this is an uncorrelated statistical process 
with the same $f$ for each proton. What is the value of $p$, so that 
$\delta t$ of Eq.~(\ref{deviation}) is obtained?

Under these assumptions the expectation value of $t$ becomes
\begin{eqnarray} \label{delt}
  \langle t\rangle &=& \frac{e^{-a\,t_{\max}}}{2\,t_{\max}} 
  \int_{-t_{\max}}^{t_{\max}} t\,e^{a\,t}\,dt \\ 
  &=& \frac{e^{-a\,t_{\max}}}{2\,t_{\max}} \frac{d~}{d a} 
  \int_{-t_{\max}}^{t_{\max}} e^{a\,t}\,dt \\
  &=& \frac{e^{-a\,t_{\max}}}{t_{\max}} \frac{d~}{d a}\, 
      \frac{\sinh(a\,t_{\max})}{a}\,.
\end{eqnarray}
For uniform creation probabilities $a=0$ and $\langle t\rangle=0$ 
as it should. Now, we want to produce the effect (\ref{deviation}),
which is small compared to $t_{\max}$. Therefore, the leading
Taylor expansion of $\sinh(a\,t_{\max})/a$ is sufficient
\begin{eqnarray} \label{a}
  \langle t\rangle &=& \frac{e^{-a\,t_{\max}}}{t_{\max}} 
  \frac{d~}{d a}\,\left( t_{\max} + \frac{a^2\,t_{\max}^3}{3!}
  + \dots \right)\\
  & =& \frac{a}{3}\, e^{-a\,t_{\max}}\,t_{\max}^2\ =\
  - \frac{t_{\max}^3}{3}\,a^2 + \frac{t_{\max}^2}{3}\,a + \dots\ .
\end{eqnarray}
Solving the quadratic equation with $t_{\max}= 1,473\,{\rm ns}$ 
and $\langle t\rangle = 60\,{\rm ns}$ gives two solutions for $a$
of which the physical (discuss why) one is $a=9.67\times 10^{-5}\,
{\rm ns}^{-1}$. 

As we assume our protons uniformly distributed and there are $6.2\times 
10^{15}$ protons in one extraction, the time interval associated with
one proton is 
\begin{eqnarray} \label{tP}
  t_P=\frac{t_{\max}}{6.2\times 10^{15}} = 2.37\times 10^{-13}\,{\rm ns}\
\end{eqnarray}
and the factor $f$ of Eq.~(\ref{f}) becomes
\begin{eqnarray} 
  f\ =\ \exp\left(-a\,t_P\right)\ =\ 1 - p
\end{eqnarray}
with
\begin{eqnarray} \label{p}
  p\ =\ a\,t_P\ =\ 1.38\times 10^{-16}\,.
\end{eqnarray}

{\bf Discussion:}
Is it obvious that $0<p\ll 1.4\times 10^{-16}$ holds in the CNGS 
experiment?  As described in detail by the CNGS collaboration 
\cite{CNGS}, the neutrino extraction succeeds in a cascade of 
events and the present authors understand nothing about the 
experimental details and whether any parts of the process could 
be affected by initial protons. The purpose of our simple 
exercises can only be to provide a starting point for physics 
discussions with university undergraduates or students in gifted 
classes at high schools. Our idea is to address some of the issues 
of the experiment while staying within the calculus skills of our 
target group.
Continuing our calculation within these limitations, the probability 
of the last proton to produce a neutrino is $f^n=\exp(-a\,t_{\max})=
86.7\%$ ($n=6.2\times 10^{15}$) of the corresponding probability of 
the first proton. This number is visually too small when compared
with the neutrino arrival time distribution of Fig.~11 of \cite{CNGS}.

Students may want to vary the parameters used. For instance, if one 
takes $t_{\max}=5,000\,{\rm ns}$ as suggested by the quoted figures, 
assumes $\triangle t = 22.5\,{\rm ns}$ and takes out two standard 
deviations (\ref{eb2_t}) from the signal (\ref{deviation}), one has only 
to explain $\delta t = 15\,{\rm ns}$. Crunching the numbers again one 
ends up with an efficiency of $f^n = 99.1\%$ for the last proton when 
compared to the first, which is no longer in visual discrepancy with
Fig.~11. A possible modification of our approach is to let $p$ depend 
on the proton number $i=1,\dots,n$, $p(1)>p$ ($p$ as calculated before) 
and $p(i)\to 0$ with increasing $i$, a model suitable for computer 
simulations.

Finally, one could ask the students to design an experiment that would 
efficiently rule out a bias in the departure times. A solution would 
be to install a neutrino detector in the mountains close to CERN, 
because $\delta t$ does then not depend on the distance of the 
detector. Besides, critical students will presumably invent all 
kind of potential error sources, which can be ruled out by careful
analysis. We hope that our exercises will serve educators and their 
students well to have a fun class or two about a hot topic.
\medskip

\acknowledgments Berg is in part supported by the DOE grant 
DE-FG02-97ER41022, Hoeflich by NSF grant 10089620708855.

\end{document}